\documentclass[conference]{IEEEtran}
\IEEEoverridecommandlockouts
\usepackage{amsmath,amssymb,amsfonts}
\usepackage{algorithmic}
\usepackage{graphicx}
\usepackage{textcomp}
\usepackage{xcolor}
\usepackage{graphicx}  
\usepackage{etex}
\usepackage{cite}
\usepackage{svg}      
\usepackage{caption}   
\usepackage{array} 
\usepackage{stfloats}
\usepackage{tabularx}
\usepackage{amsmath}
\usepackage{changepage}
\usepackage[hidelinks]{hyperref}
\usepackage{float}
\usepackage{nomencl}
\def\BibTeX{{\rm B\kern-.05em{\sc i\kern-.025em b}\kern-.08em
    T\kern-.1667em\lower.7ex\hbox{E}\kern-.125emX}}
\newcommand{\nomenentry}[2]{%
  \noindent\hangindent=0.3em\hangafter=0%
  \makebox[3.9em][l]{\hspace{0.2em}$#1$}
  #2\par\vspace{0.6ex}
}
\begin{document}

\title{System-Level Impacts of Flexible Data Center Load Scheduling on Cost, Emissions, and Transmission Congestion\\
}

\author{
\IEEEauthorblockN{Akibul Hasan Mazumder and Yuanrui Sang}
\IEEEauthorblockA{\textit{Department of Electrical and Computer Engineering} \\
\textit{University of Massachusetts Amherst}\\
Amherst, USA \\
amazumder@umass.edu, ysang@umass.edu}
}

\maketitle
\begin{abstract}
Large data centers are being deployed in the U.S. at an unprecedented rate, introducing significant flexible load potential. A portion of data center workloads—best-effort (BE) jobs—can be scheduled flexibly to reduce power system operating costs and emissions. However, the system-level impacts of such scheduling remain underexplored. This paper investigates the effects of flexible data center load scheduling on operating cost, system stress, and emissions using the \textit{ACTIVSg2000} 2000-bus test system. Results show that BE loads shift toward periods of lower locational marginal prices (LMPs), typically aligned with high renewable generation. Importantly, latency-critical (LC) workloads remain unaffected, preserving quality of service (QoS). Flexible scheduling also leads to reductions in both greenhouse gas and toxic emissions, as well as transmission congestion, compared to inflexible operation, demonstrating its potential to support more efficient and sustainable grid operation.
\end{abstract}


\section*{Nomenclature}

\textbf{Sets/Indices} \\

\nomenentry{\mathbb{B}}{Set of buses}%
\nomenentry{\mathbb{G}}{Set of generators}%
\nomenentry{\mathbb{G}_i}{Set of generators connected to bus $i$}%
\nomenentry{\mathbb{L}}{Set of transmission lines}%
\nomenentry{\mathbb{S}}{Set of linearized generation segments}%
\nomenentry{\mathbb{T}}{Set of dispatch intervals}%
\nomenentry{\mathbb{\phi}^{in}_i}{Set of lines entering bus i}%
\nomenentry{\mathbb{\phi}^{out}_i}{Set of lines leaving bus i}%

\vspace{1ex}
\textbf{Parameters} \\

\nomenentry{b_\ell}{Susceptance of line $\ell$}%
\nomenentry{C^{nl}_g}{No load cost of generator $g$}%
\nomenentry{C^{seg}_{g,s}}{Cost coefficient of segment $s$ of generator $g$}%
\nomenentry{F^{min}_\ell}{Lower bound of power flow of line $\ell$}%
\nomenentry{F^{max}_\ell}{Upper bound of power flow of line $\ell$}%
\nomenentry{GHG_g}{
\parbox[t]{\dimexpr\linewidth-\nomlabelwidth\relax}{
$CO_2$ equivalent GHG emission rate (lbs/MWh) of\\
the generator $g$
}
}
\nomenentry{P^{max}_{DC_i}}{Maximum power rating of data center at bus $i$}%
\nomenentry{P^{seg}_{g,s}}{Upper limit of segment $s$ of generator $g$}%
\nomenentry{R^{down}_g}{Ramp down limit of generator $g$}%
\nomenentry{R^{up}_g}{Ramp up limit of generator $g$}%
\nomenentry{S_{base}}{Base MVA of the system}%
\nomenentry{tox_g}{%
\parbox[t]{\dimexpr\linewidth-\nomlabelwidth\relax}{%
Toluene equivalent toxic emission rate (lbs/MWh)\\
of the generator $g$
}%
}
\nomenentry{\Delta t}{Time length of one dispatch interval}%
\nomenentry{\theta^{min}_i}{Minimum voltage angle of bus $i$}%
\nomenentry{\theta^{max}_i}{Maximum voltage angle of bus $i$}%

\vspace{1ex}
\textbf{Variables} \\

\nomenentry{D_{i,t}}{Load demand at bus $i$ at time $t$}%
\nomenentry{E^{BE}_i}{MWh requirement of BE tasks at bus $i$}%
\nomenentry{F_{\ell,t}}{Active power flow of line $\ell$ at time $t$}%
\nomenentry{LMP_t}{Vector containing LMPs at all buses at time $t$}%
% \nomenentry{$P^{seg}_{g,s,t}$}{%
% \parbox[t]{\dimexpr\linewidth-\nomlabelwidth\relax}{%
% Power generation of segment $s$ of generator $g$ at\\time $t$%
% }%
% }
\nomenentry{P^{seg}_{g,s,t}}{Power generation of segment $s$ of generator $g$ at\\ \indent ~~~~~~~~~time $t$}%

\nomenentry{P_{g,t}}{Total generation by generator $g$ at time $t$}%
\nomenentry{P^{aux}_{i,t}}{%
\parbox[t]{\dimexpr\linewidth-\nomlabelwidth\relax}{%
Auxiliary load (all other loads except BE and\\
LC loads) of data centers at bus $i$ at time $t$%
}%
}
\nomenentry{P^{BE}_{i,t}}{BE load at bus $i$ at time $t$}%
\nomenentry{P^{LC}_{i,t}}{LC load at bus $i$ at time $t$}%
\nomenentry{\theta_{i,t}}{Voltage angle of bus $i$ at time $t$}%

\section{Introduction}

The electric power consumption of data centers is increasing at an accelerated rate. Between 2014 and 2018, the compound annual growth rate was 7\%, which then increased to 18\% between 2018 and 2023, and is expected to have a gigantic leap of up to 27\% between 2023 and 2028 \cite{1}. Naturally, this leads to a question: is this dramatic rise in electric power consumption by data centers only a burden on the already congested transmission grids, or is there a way to utilize those huge loads for more efficient power consumption by providing some form of service to the power system? The answer is yes, as shown in numerous studies.

Many studies have explored the possibility of data centers providing grid services. Jahanshahi \textit{et al.} \cite{2} introduces PowerMorph, a quality of service (QoS)-aware data center power resahping framework that enables data centers to effectively participate in frequency regulation service, ensuring low cost for power consumption. In this study, best-effort (BE) tasks are colocated as a 'complementary workload' with the latency-critical (LC) tasks to ensure high provisioning capacity. Consequently, the BE tasks' energy consumption characteristics allow a portion of the total data center power consumption to exhibit a certain degree of flexibility, as discussed in \cite{3}. Wang \textit{et al.} \cite{4} develops a comprehensive framework for data center frequency regulation service provision in both hour-ahead market and real-time operations using a risk-constrained hour-ahead bidding strategy and a control mechanism for real-time power consumption, respectively, in order to minimize electricity costs and response time to frequency regulation signal. By designing a bi-layer hierarchical control scheme for optimal segregation of the regulation signal between the DC and UPS batteries, Kaur \textit{et al.} \cite{5} suggest the use of data centers and UPS units to achieve effective frequency regulation. In \cite{6}, workload modulation and UPS coordination options for data centers are explored to provide fast frequency response, which is demonstrated using a modified IEEE 39-bus system. Kez \textit{et al.} \cite{7} explores utilizing UPS storage systems, HVAC cooling infrastructure, and the capability to temporarily disconnect the entire facility from the grid during transient frequency disturbances to provide fast frequency response. They measure the success of their approaches using frequency metrics, such as the frequency nadir and the rate of change of frequency (RoCoF) of the system during transients. By adhering to device operating conditions and constraints, a framework for cooperative participation of data centers' best-effort workloads and backup power distribution units to provide fast frequency response is introduced in \cite{8}. Fu \textit{et al.} \cite{9} proposed a control strategy for frequency regulation using both IT and cooling systems of data centers. Adopting chance-constrained programming in order to enhance the participation of data centers in frequency regulation markets, Zou \textit{et al.} \cite{10} proposes a coordinated bidding strategy for electricity buying and regulation capacity offerings of multiple data centers that are distributed geographically. An ancillary service model (ASM) for data centers and power systems is discussed in \cite{11}, where the authors also propose a service-level agreement (SLA) between data centers and power systems to provide mutual benefits.

The rapid expansion of data centers fueled by AI and cloud computing has made them one of the largest consumers of electricity, necessitating the study of their electricity consumption within advanced power system analysis frameworks to ensure their economic and reliable operation \cite{12}. These large, geographically distributed loads are gradually changing the traditional load patterns of the power systems to which they are connected, necessitating a more critical study of concepts such as marginal pricing, network congestion, and dispatch strategies. Also, there is scope to explore the indirect environmental impacts (indirect because their power consumption might employ more renewable and/or fossil fuel generation) of data centers considering the flexible portion of their power consumption, since traditional economic dispatch will try to schedule the flexible loads to the periods of low locational marginal pricing, usually associated with high renewable generation \cite{13}.

Prior work, such as \cite{14}, performed economic dispatch on the modified \textit{IEEE RTS-96} 24-bus system by connecting a 20 MW data center at different locations across case studies and comparing scenarios with and without flexibly scheduled data center load. Although this provides insights into important power system quantities such as optimized generation cost and locational marginal prices under flexible and fixed scheduling of data center loads, the test system lacks resemblance with real systems since data centers in practice are more likely to be connected to much larger systems, with a more complex arrangement of the system resources. No study exists, to the best of our knowledge, that optimizes generation costs through flexible scheduling of multiple data centers connected to a large power system. A lack of emission data for large systems is also observed. Hence, investigation of emission trends in large power systems with high penetration of data center loads is notably absent. To fill this gap, this study explores the optimal scheduling of BE workloads across a large number of data centers distributed across various locations within a large power system, subject to the system's physical constraints. It also focuses on the system's emission characteristics, using emission parameters for the fossil-fuel generators derived from publicly available emission data.

The main contributions of the paper are as follows:
\begin{itemize}
     % \item It divides each data center load into two portions -- one portion has the ability to be scheduled flexibly across the whole dispatch period, the other portion does not. In this way, a resemblance with practical data center power consumption is maintained, as data centers have both BE and LC tasks \cite{2}.
     % \item The data centers' energy consumption model is integrated into the physical constraints of economic dispatch model of generators in power systems.
     \item Optimal data center scheduling is conducted in a large-scale power system, the \textit{ACTIVSg2000} 2000-bus system, with realistic data center deployment.
     \item The impacts of flexible data center scheduling on large-scale power system operating costs, transmission congestion, and emissions are studied.
     \item Emission rate data for generators, along with data center locations and capacities, are incorporated into the \textit{ACTIVSg2000} system. The updated system data is available at \cite{18} and can be used by the research community for future studies.
\end{itemize}

The rest of the paper is organized as follows.
\hyperref[sec:model]{Section~II} discusses the model formulation.
\hyperref[sec:setup]{Section~III} focuses on the simulation setup.
\hyperref[sec:analysis]{Section~IV} describes some case studies and analyzes the results observed in the case studies.
Lastly, \hyperref[sec:conclusion]{Section~V} concludes the paper by providing key insights and suggesting some possible future work.

\section{Model Formulation}

The best-effort (BE) tasks of the data centers are less stringent in terms of time criticality than the latency-critical (LC) tasks \cite{2}. This work assumes that quality of service (QoS) of data centers is maintained as long as the BE tasks are performed within the 24-hour economic dispatch window, whereas the LC tasks must be completed as soon as possible, ideally, instantly. Hence, it is reasonable to assume that there is a definite amount of energy consumption requirement for each data center related to its BE tasks, and it is possible to schedule this portion of power across the intervals of the economic dispatch framework as long as that energy consumption requirement is met. That being said, this work divides the power consumption of each data center into three parts: (1) power related to the BE tasks (can be flexibly scheduled), (2) power related to the LC tasks (cannot be flexibly scheduled), and (3) auxiliary power consumption, such as network equipment, power conditioning equipment, cooling, air conditioning, lighting, etc.

The data center power consumption model described above is integrated into an hourly economic dispatch framework over a 24-hour window represented by \eqref{1}-\eqref{12}. Equation \eqref{1} represents the optimization objective function, which is essentially the minimization of the total generation cost by considering no-load generating costs and piecewise-linear generating costs of the conventional fossil fuel and the hydro generators (nuclear generators are considered base-load generators, and hence non-dispatchable; renewables except hydro are also considered non-dispatchable). Equations \eqref{2}-\eqref{12} represent the power system's physical constraints. The sum of the segment-wise generations is the total generation by a generator, as shown in \eqref{2}. Equation \eqref{3} ensures that the segment-wise generation of each generator is bounded by the lower and the upper limits. The limits on ramping up and ramping down of generators between adjacent dispatch periods are enforced by \eqref{4} and \eqref{5}, respectively. Equation \eqref{6} makes sure that the transmission line active power flows are within the thermal limits of each line. The linearized DC power flow formula considers the phase angles (in radians) of the two end buses of each line, as shown in \eqref{7}. The slack bus (bus 1) voltage angle is set at 0 radian by \eqref{8}. Since this work is based on DC optimal power flow (DCOPF), \eqref{9} ensures that the bus voltage angles are within reasonably narrow limits, to enforce the linearization assumption. Nodal power balance, i.e., power flowing into the bus must equal power flowing out, is shown in \eqref{10}, where there are three terms related to data center power consumption on the right-hand side. The duals of the nodal balance constraints give the locational marginal prices (LMP) at all buses. Equation \eqref{11} makes sure that the power consumption related to the BE tasks in each dispatch interval multiplied by the length of the corresponding dispatch interval equals the total energy consumption requirement of the BE tasks at each bus. Lastly, \eqref{12} enforces that the total power consumption of each data center must be within its peak power consumption.

\begin{equation}
min\sum_{t \in \mathbb{T}}\sum_{g \in \mathbb{G}} \left(C^{nl}_g+\sum_{s \in \mathbb{S}} C^{seg}_{g,s} \cdot P^{seg}_{g,s,t}\right) \cdot \Delta t \label{1}
\end{equation}
\begin{equation}
P_{g,t} = \sum_{s \in \mathbb{S}} P^{seg}_{g,s,t}, \forall t \in \mathbb{T}, \forall g \in \mathbb{G} \label{2}
\end{equation}
\begin{equation}
0 \le P^{seg}_{g,s,t}\le P^{seg}_{g,s}, \forall g \in \mathbb{G}, \forall s \in \mathbb{S}, \forall t \in \mathbb{T} \label{3}
\end{equation}
\begin{equation}
P_{g,t}-P_{g,t-1} \le R^{up}_g, \forall g \in \mathbb{G}, \forall t \in \mathbb{T}-\{1\} \label{4}
\end{equation}
\begin{equation}
P_{g,t-1}-P_{g,t} \le R^{down}_g, \forall g \in \mathbb{G}, \forall t \in \mathbb{T}-\{1\} \label{5}
\end{equation}
\begin{equation}
F^{min}_\ell \le F_{\ell,t} \le F^{max}_\ell, \forall \ell \in \mathbb{L}, \forall t \in \mathbb{T} \label{6}
\end{equation}
\begin{equation}
F_{\ell,t} = -b_\ell \cdot S_{base} \cdot \left(\theta_{i,t}-\theta_{j,t}\right), \forall \ell \in \mathbb{L}, \forall t \in \mathbb{T}, i,j \in \mathbb{B} \label{7}
\end{equation}
\begin{equation}
\theta_{1,t} = 0,\forall t \in \mathbb{T} \label{8}
\end{equation}
\begin{equation}
\theta^{min}_i \le \theta_{i,t} \le \theta^{max}_i, \forall i \in \mathbb{B}-\{1\}, \forall t \in \mathbb{T} \label{9}
\end{equation}
%\begin{equation}
\begin{align}
\sum_{g \in \mathbb{G}_i}P_{g,t}+\sum_{k \in \mathbb{\phi}^{in}_i}F_{k,t}-\sum_{m \in \mathbb{\phi}^{out}_i}F_{m,t} = D_{i,t} \nonumber \\ +P^{LC}_{i,t}+P^{BE}_{i,t}+P^{aux}_{i,t}, \forall i \in \mathbb{B}, \forall t \in \mathbb{T} \label{10}
\end{align}
%\end{equation}
\begin{equation}
\sum_{t \in \mathbb{T}}P^{BE}_{i,t} \cdot \Delta t = E^{BE}_i, \forall i \in \mathbb{B} \label{11}
\end{equation}
\begin{equation}
0 \le P^{BE}_{i,t}+P^{LC}_{i,t}+P^{aux}_{i,t} \le P^{max}_{DC_i}, \forall i \in \mathbb{B} \label{12}
\end{equation}

\section{Simulation setup}

\subsection{Test System}
For implementation of the economic dispatch framework on a sufficiently large and complex system in order to resemble practical scenarios, this work chooses \textit{ACTIVSg2000} test system \cite{16}\cite{17}, which consists of 2000 buses, 544 generators (431 of them are considered dispatchable), and 3206 transmission lines. Though the buses are numbered differently in the actual system data, this paper considers numbering them starting from 1 to 2000 for ease of analysis. The economic dispatch framework is considered for 1 hour interval over a period of 24 hours; hence, there are 24 dispatch intervals. For simplicity, the generator cost model is linearized into four segments for each generator, in contrast to the original quadratic model of the generation costs in the system \cite{17}. In addition, the emission rates of greenhouse and toxic gases are developed for the \textit{ACTIVSg2000} system, based on the publicly available emission data of US power plants \cite{15}, and are made available at \cite{18}. 

\subsection{Placement of Data Centers}
47 data centers with varying peak power consumption were deployed across 47 buses. The power ratings of the data centers were obtained from \cite{19}. The locations of the data centers were selected by comparing the maps in \cite{19} with the PowerWorld binary file of the \textit{ACTIVSg2000} 2000-bus system, using visual inspection. It is important to note that the data center locations might not be exact representations of the actual locations, since the test system was synthetically created using statistical interference on the footprint of the Texas grid \cite{16}. The data center information for the \textit{ACTIVSg2000} system used in this study is available at \cite{18}.

\subsection{Clustering the Data Centers into Disjoint Sets}
The data centers were clustered into three mutually exclusive sets according to their locations in the system. The clustering was done in such a way that the total power ratings of the data centers in each of the clusters are almost the same. Following prior work that uses LMP dispersion as a proxy for congestion severity (e.g., \cite{20}, \cite{21}), we define a \textit{congestion metric} as the temporal average of the cross-sectional variance of nodal LMPs:
\begin{equation}
\gamma = \frac{1}{\lvert \mathbb{T} \rvert} \sum_{t \in \mathbb{T}} var\left(LMP_{t}\right)
\end{equation}

The clusters are used for case studies and analysis. The details of the clusters are listed in Table~\ref{tab1}.
\begin{table}[htbp]
\caption{Cluster Data}
\begin{center}
\begin{tabular}{|c|c|c|c|c|}
\hline
\textbf{}&\textbf{Whole System}&\textbf{Cluster 1}&\textbf{Cluster 2}&\textbf{Cluster 3} \\
\hline
\textbf{Bus No.} & 1--2000 & 1--586 & 587--1168 & 1169--2000 \\
\hline
\textbf{Total DC} & 9047.61 & 2870.95 & 3061.61 & 3115.05 \\
\textbf{Cap. (MW)} & & & &\\
\hline
$\boldsymbol{\gamma}$ \textbf{(\$/MWh)}& 0.4961 & 1.3473 & 0.0715 & 0.1142 \\
\hline
\end{tabular}
\label{tab1}
\end{center}
\end{table}

\subsection{Load Categorization of Data Centers}
Data centers comprise multiple power-consuming components, including servers, cooling systems, and networking equipment. In this study, 60\% of each data center's instantaneous power consumption is attributed to servers. $30 \pm 10\%$ of the total server consumption is designated as LC tasks' consumption. The remaining server power is associated with the BE tasks, and multiplying it by the length of the entire dispatch period yields the BE tasks' daily energy consumption. In this study, the term \textit{fixed load} refers to the case when the power related to the BE tasks is evenly distributed among each hour of the 24-hour dispatch period, whereas \textit{flexible load} indicates the case where flexible scheduling of the BE power consumption is implemented. The power not consumed by BE tasks is considered non-schedulable. 

\section{Case Studies and Result Analysis}

\subsection{Flexible Scheduling across Time}

\begin{figure}[htbp]
\centerline{\includegraphics[width=0.8\columnwidth]{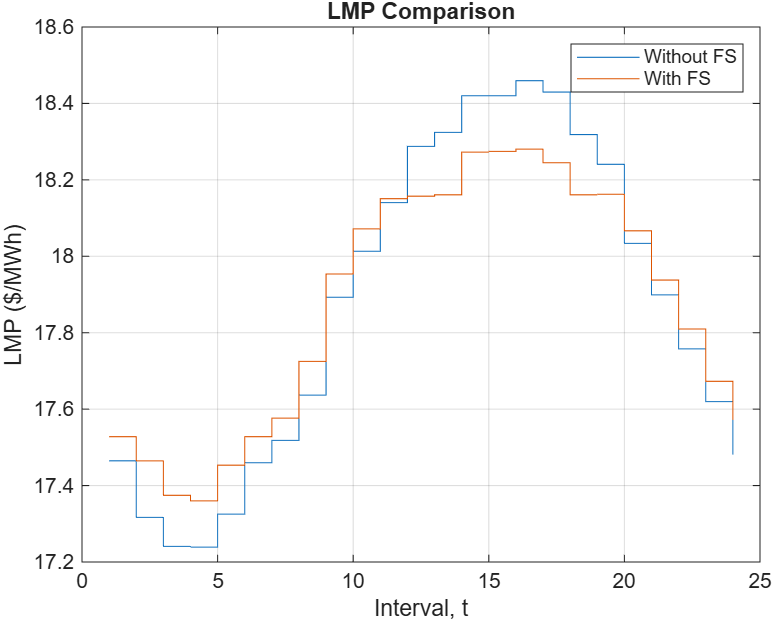}}
\caption{Comparison of LMPs at bus no. 1995}
\label{fig1}
\end{figure}

\begin{figure}[htbp]
\centerline{\includegraphics[width=0.8\columnwidth]{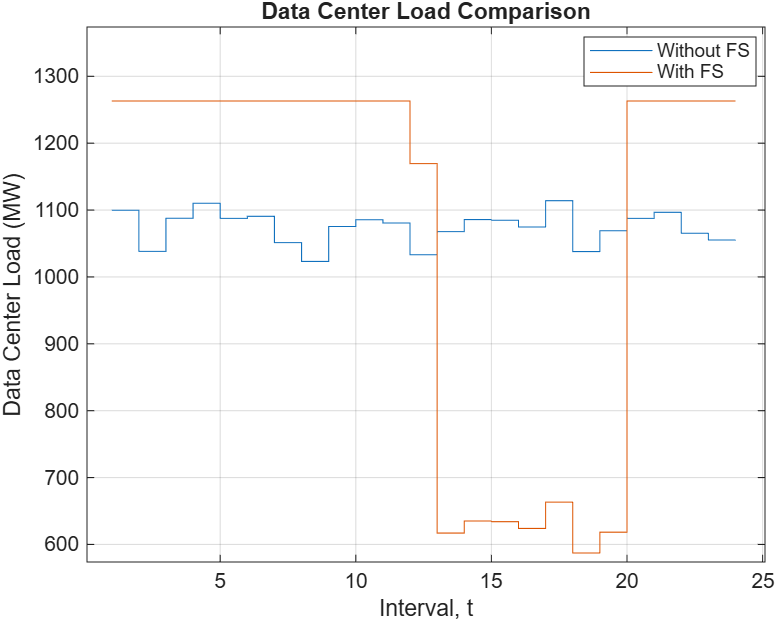}}
\caption{Comparison of hourly load distribution of the data center at bus no. 1995}
\label{fig2}
\end{figure}

\begin{figure}[htbp]
\centerline{\includegraphics[width=0.8\columnwidth]{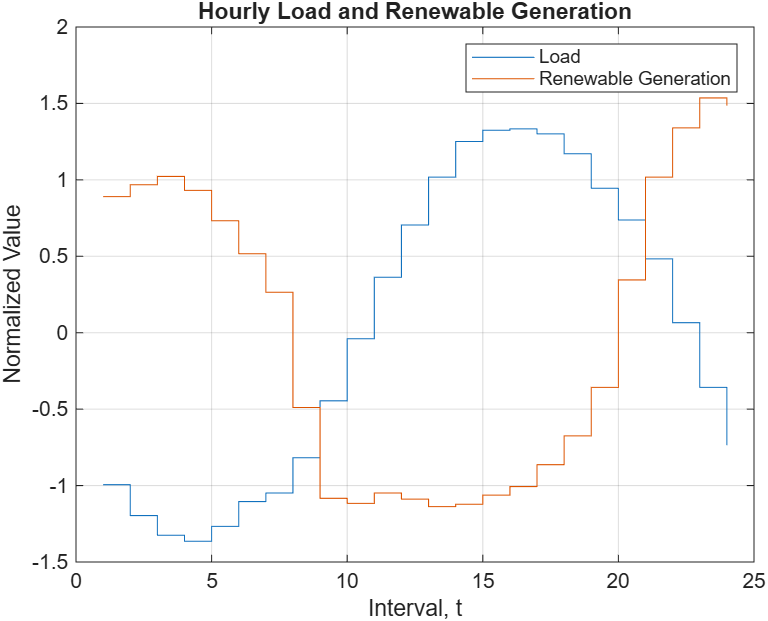}}
\caption{Temporal distribution of load demand (except data center loads) and renewable generation}
\label{fig3}
\end{figure}

This subsection analyzes the locational marginal prices (LMP) and the distribution of the flexibly scheduled loads. Fig.~\ref{fig1} compares the LMPs on bus 1995 between the cases with flexible scheduling (FS) in the whole system and without FS, where the data center with the highest power rating (1263 MW) is located. Although in off-peak hours LMPs are slightly higher with FS, FS helps to reduce the LMPs during peak hours significantly. Fig.~\ref{fig2} provides a visual contrast between FS and without FS data center loads at bus 1995. It is observed that the FS framework helps maintain comparatively lower demand during peak hours, which is remarkable. Without the aid of any external regulation signal, the economic dispatch framework discussed in this work can schedule the flexible portions of the loads to periods of comparatively lower electricity prices associated with high renewable generation, considering physical system constraints, while maintaining the total energy consumption requirement described by \eqref{11}. Total load demand, excluding data center loads, and total renewable generation are shown in Fig.~\ref{fig3} to correlate LMPs with off-peak hours and high renewable generation availability.


\subsection{Generation Costs for Flexible and Fixed Scenarios}
\begin{figure}[htbp]
\centerline{\includegraphics[width=0.8\columnwidth]{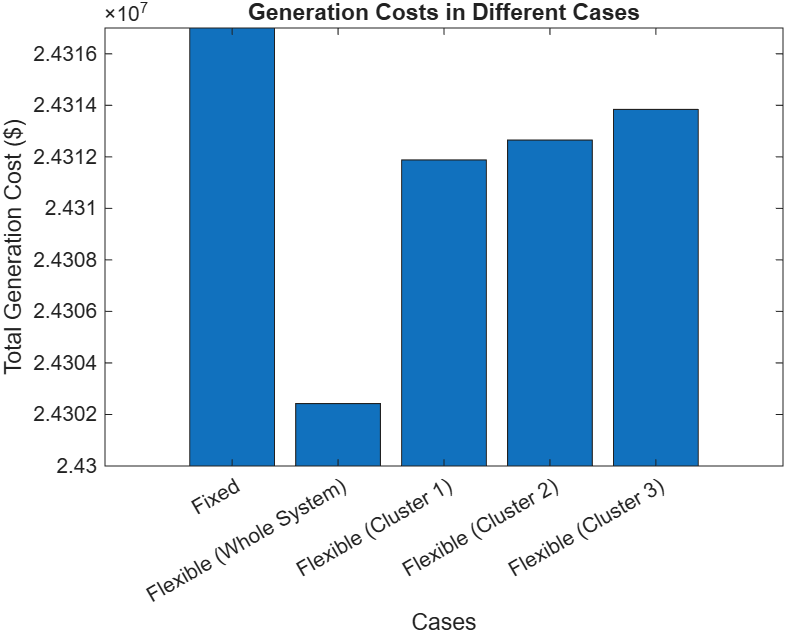}}
\caption{Comparing generation costs for different simulation cases}
\label{fig4}
\end{figure}
Comparison of generation costs among different case studies is performed in this subsection. As shown in Fig.~\ref{fig4}, the total electricity generation cost is the highest when there is no flexible scheduling, the amount is around \$24,318,753.28. The case when all the data centers are under flexible scheduling consideration gives the lowest cost (approximately \$24,302,424.76). In the cases when only one of clusters 1, 2, and 3 are flexibly scheduled while keeping the remaining portion of the system in fixed load condition, the generating costs are around \$24,311,879.43, \$24,312,650.73, and \$24,313,839.73, respectively, which are lower than the case when the whole system was in fixed load condition, but higher than the one when all of the data centers were flexibly scheduled. However, no correlation between the \textit{congestion metric} of the clusters and the generation cost was observed.

\subsection{Stress on Transmission Lines}

\begin{figure}[h]
\centerline{\includegraphics[width=0.8\columnwidth]{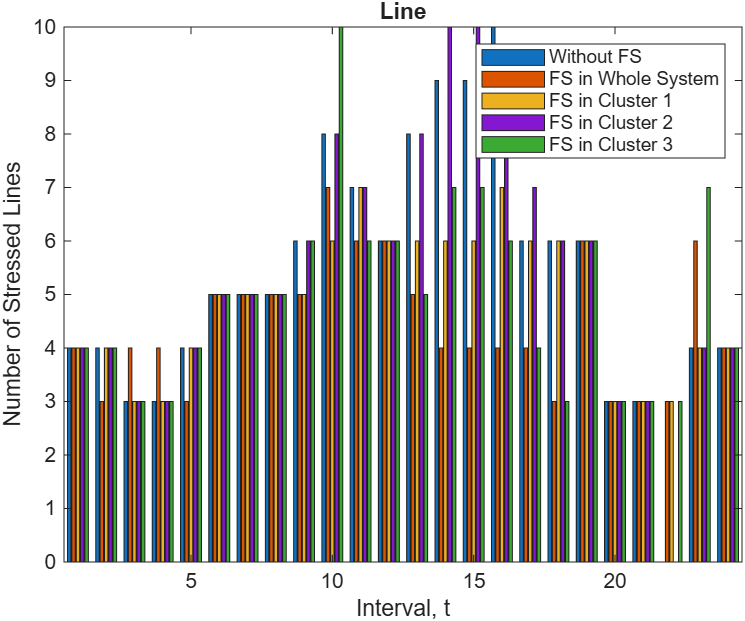}}
\caption{Stressed Transmission Lines in Different Cases}
\label{fig5}
\end{figure}

Fig.~\ref{fig5} compares the number of stressed lines in the system across all the dispatch intervals in all of the 5 cases. The total number of congested lines in all the intervals in each case is listed in Table~\ref{tab2}. Lines are considered stressed if and only if they are operating at 90\% or more of their respective active power ratings. Table~\ref{tab2} shows that FS across the whole system resulted in the fewest congested lines. In the cluster cases, FS in clusters with higher congestion metrics showed good performance, due to the number of stressed lines; the case with FS in cluster 2 (its congestion metric is negligible compared to that of the others) did not impact the number of stressed lines significantly compared to the case without FS. In sum, FS tends to reduce transmission system stress more effectively in more congested areas.

\begin{table}[htbp]
\caption{Total Number of Stressed Lines}
\begin{center}
\begin{tabular}{|c|c|c|c|c|}
\hline
\textbf{Without FS}&\textbf{FS in}&\textbf{FS in }&\textbf{FS in}&\textbf{ FS in} \\
\textbf{}&\textbf{Whole System}&\textbf{Cluster 1}&\textbf{Cluster 2}&\textbf{Cluster 3} \\
\hline
128 & 106 & 117 & 130 & 119 \\
\hline
\end{tabular}
\label{tab2}
\end{center}
\end{table}

\subsection{Greenhouse Gas Emission}

\begin{figure}[h]
\centerline{\includegraphics[width=0.8\columnwidth]{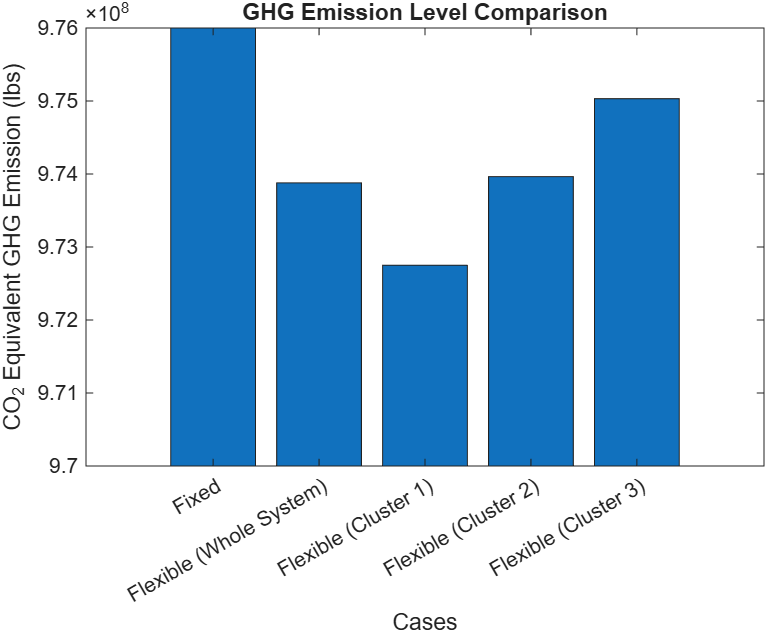}}
\caption{GHG Emission in Different Cases}
\label{fig6}
\end{figure}

For all the simulation case studies, using the data for $CO_2$ equivalent output greenhouse gas (GHG) emission rate (lbs/MWh) for each of the generators in the system, the total emissions in lbs were calculated using the following equation:
\begin{equation}
Total\,GHG\,Emission = \sum_{g \in \mathbb{G}} \sum_{t \in \mathbb{T}} P_{g,t} \cdot GHG_g \cdot \Delta t
\end{equation}
Fig.~\ref{fig6} shows the comparison among the GHG emission levels in each case. All the flexible load cases outperform the fixed load case in reducing GHG emissions, a remarkable feat since no explicit optimization criterion for minimizing GHG emissions is used in the objective function of the economic dispatch framework.

\subsection{Toxic Gas Emission}

\begin{figure}[h]
\centerline{\includegraphics[width=0.8\columnwidth]{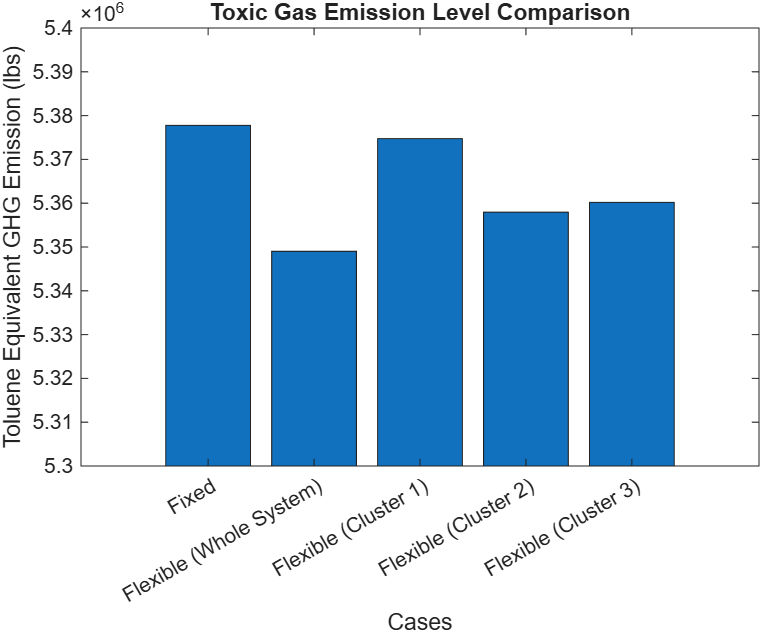}}
\caption{Toxic Emission in Different Cases}
\label{fig7}
\end{figure}

For the \textit{ACTIVSg2000} system, $NO_x$ and $SO_2$ emission rate (lbs/MWh) for each of the generators were created by a noisy data fitting technique using publicly available toxic gas emission data from \cite{15}. In each of the case studies, total toxic gas emission is calculated by:
\begin{equation}
Total\,Toxic\,Emission = \sum_{g \in \mathbb{G}} \sum_{t \in \mathbb{T}} P_{g,t} \cdot tox_g \cdot \Delta t
\end{equation}
The HTP factors for converting into toluene equivalent emission were obtained from \cite{22}. The emission level comparison in different test cases are shown in Fig.~\ref{fig7}. As with the GHG emission, this study finds that flexible scheduling provides a lower level of toxic gas emission as a by-product of cost optimization in the economic dispatch framework, a win-win situation.

\section{Conclusions}

This work studies the impacts of flexible scheduling of BE workloads in the data center on a large power system. It shows that flexible scheduling helps reduce system generating costs by utilizing low-cost electricity periods for economic operation, without the need for any regulatory signal. Integrating the requirements of flexible BE workloads within the physical system constraints of power system operations, this study also finds that this type of scheduling helps reduce environmental pollution, including greenhouse gas and toxic gas emissions. All of these benefits can be realized without compromising the stringent, time-sensitive requirements of LC workloads in data centers. It is also found that this flexible scheduling approach tends to reduce transmission congestion, which is a positive outcome, given that transmission congestion has become a bottleneck for data center integration into the power grid.

\bibliographystyle{IEEEtran}
\bibliography{Ref.bib}

\end{document}